\newif\iffinal
\finalfalse
\iffinal
\documentclass[10pt,journal,compsoc]{IEEEtran} %
\else
\documentclass[11pt,onecolumn,draftcls]{IEEEtran} %
\fi

\usepackage{multirow}
\usepackage{amsfonts,amssymb,amsmath,amsthm,mathrsfs}
\usepackage{algorithmic,algorithm}
\usepackage{xcolor,soul}
\usepackage{graphicx}
\usepackage{ifpdf}
\ifpdf
\usepackage[]{graphicx}
\else
\fi
\usepackage{array}
\usepackage{caption}
\usepackage{subcaption}
\usepackage{url,cite}
\usepackage{breakurl}

\newtheorem{definition}{Definition}

\newtheorem{lemma}{Lemma}

\newtheorem{remark}{Remark}
\newtheorem{theorem}{Theorem}
\usepackage{setspace}

\usepackage[normalem]{ulem}
\definecolor{revision}{rgb}{1,0,0}

\def\mbfb{\mathbf{b}}

\def\mbfx{\mathbf{x}}
\def\mbfy{\mathbf{y}}
\def\idi{\mathsf{ID}_i}

\begin{document}
\title{PassBio: Privacy-Preserving User-Centric Biometric Authentication}

\author{Kai Zhou and Jian Ren\thanks{The authors are with the Department of Electrical and Computer Engineering,Michigan State University, East Lansing, MI 48824-1226, Email: \{zhoukai, renjian\}@msu.edu}}

\maketitle

\begin{abstract}
The proliferation of online biometric authentication has necessitated security requirements of biometric templates. The existing secure biometric authentication schemes feature a \emph{server-centric} model, where a service provider maintains a biometric database and is fully responsible for the security of the templates. The end-users have to fully trust the server in storing, processing and managing their private templates. As a result, the end-users' templates could be compromised by outside attackers or even the service provider itself. In this paper, we propose a \emph{user-centric} biometric authentication scheme (PassBio) that enables end-users to encrypt their own templates with our proposed light-weighted encryption scheme. During authentication, all the templates remain encrypted such that the server will never see them directly. However, the server is able to determine whether the distance of two encrypted templates is within a pre-defined threshold.
 Our security analysis shows that no critical information of the templates can be revealed under both passive and active attacks. PassBio follows a ``compute-then-compare'' computational model over encrypted data. More specifically, our proposed Threshold Predicate Encryption (TPE) scheme can encrypt two vectors $\mbfx$ and $\mbfy$ in such a manner that the inner product of $\mbfx$ and $\mbfy$ can be evaluated and compared to a pre-defined threshold. TPE guarantees that only the comparison result is revealed and no key information about $\mbfx$ and $\mbfy$ can be learned. Furthermore, we show that TPE can be utilized as a flexible building block to evaluate different distance metrics such as Hamming distance and Euclidean distance over encrypted data. Such a compute-then-compare computational model, enabled by TPE, can be widely applied in many interesting applications such as searching over encrypted data while ensuring data security and privacy.
\end{abstract}


\section{Introduction}

Biometric authentication has been incredibly useful in services such
as access control to authenticate individuals based on their biometric traits. Unlike passwords or
identity documents used in conventional authentication systems, biometric
traits, such as fingerprint, iris and behavioral characteristics are
physically linked to an individual that cannot be easily manipulated.
Also due to such a strong connection, security and privacy of
the biometric templates used in the authentication process is a critical
issue~\cite{jain2012biometric,jain201650,rane2013secure}. 

Existing biometric authentication systems generally employ a two-phase
mechanism \cite{rane2013secure}. In a registration phase, an end-user
submits her biometric template to the service provider who will store
the template along with the end-user's ID in a central database. In
a query phase, the end-user requesting access to certain services
will submit a fresh template to the service provider for authentication.
Based on the end-user's ID, the service provider will retrieve the
enrolled template for comparison. Only if the two templates are close
enough under certain distance metric, the end-user is successfully
authenticated.

The above biometric authentication model can be regarded as \textit{server-centric.
}That is, the service provider will receive end-users' biometric templates
in plaintext and is fully responsible for the security of the templates.
Such an approach has several inherent deficiencies. First, the end-users
have to fully trust the service provider to properly handle their
templates; otherwise the security and privacy of the templates are
at risk. For example, different service providers may crosscheck their
databases to discover possible duplications, meaning that the same
end-user may get enrolled in different services. As a consequence,
the privacy of the end-user is violated. Second, unlike password,
biometric templates are inherently noisy. As a result, the fresh template
to be authenticated is not necessarily the same as the registered
template. Such a property prevents the service provider from keeping
the templates encrypted during the whole authentication process. At
some point, the templates have to be recovered in plaintext for distance
computation and comparison. This renders the adversaries with the
opportunity to spy the registered or freshly submitted templates.

To address the above issues, we propose a \textit{user-centric} model
for biometric authentication. In terms of security, such a user-centric
model has several unique features, compared to the server-centric
model. First, biometric templates are encrypted at user side and then
transmitted to the server. The service provider is only able to see
encrypted versions of the registered templates and query templates.
Second, the secret keys and the templates are generated and processed
locally thus never leaving the local environment. Third, computations
involved in authentication are all carried out on ciphertext, meaning
that no templates are exposed in plaintext during the authentication.
These features can effectively reduce the possibility for the server
as well as outside adversaries to learn any key information of the biometric
templates. 

To meet the demands of the proposed user-centric model, the underlying
encryption scheme should be efficient and expose as little information
as possible. Since the key management and encryption are carried out
at the user side, the encryption scheme should be computationally efficient. Some
existing encryption schemes relying on heavy cryptographic operations such as  Predicate
Encryption (PE) \cite{katz2008predicate,shen2009predicate}, Inner
Product Encryption (IPE) \cite{bishop2015function,abdalla2015simple,kim2016function,datta2016functional}
and Homomorphic Encryption (HE) \cite{smart2010fully,naehrig2011can}
may not be practical in such a scenario. Also, the encryption scheme should support certain kinds of computation
on encrypted data. For example, given two encrypted vector, the server
should be able to decide whether the two vectors are close enough
(e.g., within a certain threshold) under some distance metric. The encryption scheme should expose
as little template information as possible for security and privacy. Although some distance preserving
transformation schemes \cite{wong2009secure} have been
proposed for private nearest neighbor search on encrypted data, these schemes will inevitably expose the distance information between
the registered and query template, which makes them vulnerable to
security attacks \cite{wong2009secure}.

In this paper, we propose a new primitive named Threshold Predicate Encryption (TPE). TPE  encrypts two vectors $\mbfx$ and $\mbfy$ respectively as $C_{\mbfx}$ and $C_{\mbfy}$. Unlike traditional cryptosystems, the decryption of TPE will only reveal whether the inner product of $\mbfx$ and $\mbfy$ is within a threshold $\theta$ or not, instead of the plaintext. Therefore, no more information about the vectors and the inner product are exposed. TPE is fundamentally different from  the previous schemes such as IPE~\cite{kim2016function} and PE~\cite{katz2008predicate}. IPE reveals the inner product of $\mbfx$ and $\mbfy$ thus the distance between the registered template and the query template, which makes the scheme vulnerable to security attacks \cite{wong2009secure}. PE can only reveal whether  the inner product equals to a threshold or not. It is not flexible enough for biometric authentication since generally we want to know whether the distance between the two templates is within a threshold. In comparison, our proposed TPE provides an excellent trade-off between information leakage and flexibility, which makes is uniquely suitable for biometric authentication.

TPE enables a compute-then-compare computational model over encrypted data. In this model, given ciphertexts, any party is able to \textbf{compute} the distance between the underlying plaintexts and then \textbf{compare} the distance with a threshold. The output is an indicator showing whether the distance is within the threshold or not. We show that such a computational model captures the essence of various applications such as privacy-preserving biometric identification and searching over encrypted data. TPE based schemes are able to fulfill the requirements of such applications while ensuring the security and privacy of the data.

The main contributions of this paper are summarized as follows:
\begin{itemize}
\item We propose a user-centric biometric authentication scheme enabling
end-users to utilize their biometric templates for authentication while preserving template privacy.

\item We propose a new primitive named TPE that can encrypt two
vector $\mbfx$ and $\mbfy$ in such a manner that the decryption
result only reveals whether the inner product of $\mbfx$ and
$\mbfy$ is within a threshold or not. 

\item The proposed TPE enables a compute-then-compare computational
model over encrypted data. We show that such a computational model
can be applied to many privacy-preserving applications such as biometric identification and searching over encrypted data.
\end{itemize}

The rest of this paper is organized as follows. In Section \ref{sec:Problem-Statement},
we introduce the system model as well as the threat model where different
attacks are identified. We then illustrate the design of TPE and give
a detailed implementation in Section \ref{sec:Threshold-Predicate-Encryption}.
Based on TPE, we propose the user-centric biometric authentication
scheme in Section \ref{sec:Biometric-Authentication}, where different
similarity measurements are considered. We give detailed security analysis
of TPE in Section \ref{sec:Security-Analysis}. In Section \ref{sec:Applications-of-Threshold},
we introduce some applications of TPE such as outsourced biometric
identification and searching over encrypted data. We analyze the complexity
of TPE and evaluate the performance of TPE through some simulations
in Section \ref{sec:Performance-Evaluation}. We conclude
in Section \ref{sec:Conclusion}.

\section{Related Work}

The proposed TPE scheme can be regarded as an instance of functional encryption. That is, given the decryption key, the decryption process actually produces a function of the underlying plaintext, instead of the plaintext itself. From an application point of view, biometric authentication and identification is closely related to finding the nearest neighbor of a given point (i.e.,nn or $k$-nn search). Thus, in this section, we review some related works concerning these two topics.

\subsection{Functional Encryption and Controlled Disclosure}

In conventional cryptosystem, the decryption process will eventually recover the underlying plaintext $m$. As a result, all information of $m$ is disclosed. Many applications, however, require only \emph{partially disclosure} of the information of $m$. For example, a financial organization wants to filter out those customers whose transactions exceed certain amount. For privacy concern, all the transactions of the customers are encrypted. In this case, instead of decrypting the transactions, a more desirable approach is to determine whether an transaction exceeds certain amount without disclosing the transaction. Such application scenarios motivate the research of functional encryption \cite{lewko2010fully,boneh2011functional,goldwasser2014multi}. In a functional encryption scheme, a decryption key $S_f$ is associated with a function $f$. Given the ciphertext $C$, the decryption process will evaluate the function $f(m)$, where $m$ is the underlying plaintext. Note that in this process, the plaintext $m$ cannot be recovered. Thus, by issuing different decryption keys $S_{f_i}$, functional encryption can actually implement \emph{controlled disclosure} of the plaintext $m$.

Much research effort has been devoted to designing  various functions $f_i$ for functional encryption schemes. Representative works are Predicate Encryption (PE)~\cite{katz2008predicate,shen2009predicate} and Inner Product Encryption (IPE)~\cite{bishop2015function,abdalla2015simple,kim2016function,datta2016functional}. In PE, a message is modeled as a vector $\mathbf{x}$ and a decryption key is associated with a vector $\mathbf{y}$. The decryption result is meaningful (otherwise, a random number) if and only if the inner product of $\mathbf{x}$ and $\mathbf{y}$ is equal to $0$. Based on this basic implementation, different predicates are realized such as exact threshold, polynomial evaluation and set comparison. In contract, IPE schemes will recover the value of inner product of $\mathbf{x}$ and $\mathbf{y}$, without revealing neither $\mathbf{x}$ nor $\mathbf{y}$. In the context of controlled disclosure, IPE discloses more information of the plaintext than PE. This is because with PE, one can only decide whether the inner product of $\mathbf{x}$ and $\mathbf{y}$ is equal to a certain value or not while with IPE, one can know the value of the inner product. In comparison, with TPE, what we seek is to control the amount of information to be disclosed between those of PE and IPE. As a result, TPE can efficiently fulfill the task of biometric authentication while exposing less information about the templates.

\subsection{Secure $k$-nn Search}

The problem of secure $k$-nn search can be described as finding the $k$ nearest neighbors ($k$-nn) of a given query point among a set of encrypted points. The schemes~\cite{elmehdwi2014secure,choi2014secure,wang2016practical,wong2009secure,wang2015cloudbi} for secure $k$-nn search mainly differ in the attack models they considered and the security levels they can provide. For instance, the scheme in \cite{wang2016practical} focused on search efficiency at the cost of partial privacy leakage.  Both \cite{wong2009secure} and \cite{wang2015cloudbi} considered a stronger  known-plaintext attack model. The basic ideas of these two schemes are quite similar. Given two encrypted points in the data set and one encrypted query, the comparison process in the schemes is able to determine which point is closer to the query point. Repeating this comparison process will finally reveal which point in the data set the nearest neighbor to the query point.

Our proposed TPE scheme utilizes similar techniques as that in~\cite{wang2015cloudbi}. However, the computational models as well as the security requirements are fundamentally different. In the biometric identification scheme in \cite{wang2015cloudbi}, given a query template, the server is able to identify the closest template in the database, which is returned to the end-user. After decryption of the returned template, the end-user is able to calculate the distance and determine whether the distance is within a threshold. We note that such a computational model cannot be easily applied to biometric authentication. This is because in biometric authentication, it is the server that compares the distance with a threshold while the server is not allowed to decrypt the templates thus calculating the distance. Moreover, secure $k$-nn based approaches will inherently expose more information than needed. From $k$-nn search, a sever can learn the relative distances between a query template and all the templates in the database. Such information is more than needed for biometric authentication and identification, where ideally, the server only needs to know whether the distance exceeds a pre-defined threshold.

\section{Problem Statement \label{sec:Problem-Statement}}

\subsection{System model}

We consider an online biometric authentication system consisting of two parties: an online service provider and a set of end-users. The service provider provides certain online services such as storage to its authenticated end-users. We assume that every end-user possesses a device such as a mobile phone that is able to collect the her biometric traits and transform the traits to biometric templates at the local side. Without loss of generality, we assume that each biometric template is represented by an $n$-dimensional vector $\mathbf{T}=(t_1,t_2,\dots,t_n$) of real numbers. 

\def\tci{\mathbf{T}_i=(t_{c1},t_{c2},\dots,t_{cn})}
\def\tc{\mathbf{T}_C=(t_{C,1},t_{C,2},\dots,t_{C,n})}
\def\ti{\mathbf{T}_i=(t_{i1},t_{i2},\dots,t_{in})}

The biometric authentication process consists of two phases. In the registration phase, an end-user $U_i$ will register with her biometric template $\ti$ along with a unique identifier $\idi$. We note that the template $\mathbf{T}_i$ is sent to the service provider in encrypted form denoted as $\mathsf{Enc}(\mathbf{T}_i)$ and $\idi$ can be any pseudorandom string that uniquely identifies $U_i$ within the system. The tuple $\langle\mathsf{Enc}(\mathbf{T}_i),\idi\rangle$ for the end-user $U_i$ is then stored at the server side by the service provider. In the query phase, when the end-user $U_i$ desires to authenticate herself to the service provider, $U_i$ will locally generate a fresh biometric template $\mathbf{T}'_i$ and send the tuple $\langle\mathsf{Enc}(\mathbf{T}'_i),\idi\rangle$ to the service provider, where $\mathsf{Enc}(\mathbf{T}'_i)$ is the encrypted form of $\mathbf{T}'_i$. On receiving the query, the service provider will retrieve the record $\langle\mathsf{Enc}(\mathbf{T}_i), \idi\rangle$ through searching $\idi$ in the server. Then distance between $\mathbf{T}_i$ and $\mathbf{T}'_i$ are computed based on $\mathsf{Enc}(\mathbf{T}_i)$ and $\mathsf{Enc}(\mathbf{T}'_i)$. If the distance is within a certain threshold $\theta$,  then the service provider will view the end-user $U_i$ as a valid user. We also note that during the query phase, the service provider is only able to derive whether the distance between $\mathbf{T}_i$ and $\mathbf{T}'_i$ is within the threshold $\theta$, instead of the exact distance between them.

\subsection{Threat model}

We assume the end-users are fully trusted in the registration phase. That is, they will honestly generate their own biometric templates and register at the service provider using the encrypted templates. In the query phase, we assume the encryption and decryption algorithms are publicly known. However, the secret keys are generated and kept secret at the local side throughout the whole authentication process. We do allow the adversaries to submit their own biometric templates through the local device. In this case, the local device acts as an oracle to encrypt templates and submit the encrypted templates to the service provider. The service provider can be honest-but-curious or malicious. In the former case, the service provider will honestly follow the protocol but will try to obtain any useful information of end-users' biometric templates based only on the encrypted templates. In the latter case, the adversaries may collude with the service provider such as sharing with the service provider the invalid templates that are submitted through the local devices. In summary, depending on the different capabilities of the service provider and the adversaries, we propose two attack models as follows.
\begin{enumerate}
\item Passive Attack: the service provider is able to know the registered record $\langle\mathsf{Enc}(\mathbf{T}_i), \idi\rangle$ for end-user $U_i$ and observe a series of $m$ submitted queries $\mathsf{Enc}(\mathbf{T}_i^j)$, $j=1,2,\dots,m$. However, the service provider does not know the underlying templates $\mathbf{T}_i^j$ in plaintext. Such an attack model is also known as the Ciphertext-Only-Attack in cryptography. 

\item Active Attack: besides the registered record $\langle\mathsf{Enc}(\mathbf{T}_i),\idi\rangle$ for end-user $U_i$, the service provider is able to observe a series of $m$ submitted queries $\mathsf{Enc}(\mathbf{T}_i^j)$ as well as the corresponding plaintext $\mathbf{T}_i^j$, $j=1,2,\dots,m$. Such an attack model corresponds to the Chosen-Plaintext-Attack in cryptography. In practice, an adversary may submit her own templates through the local device. The service provider can then collude with the adversary to obtain the queries in plaintext as well as the encrypted queries. 
\end{enumerate}

Informally, the security requirement of biometric authentication  is that the service provider is unable to learn any information about the templates than allowed through the authentication process. In particular, it should be possible for the service provider to determine whether the distance between two templates is within a threshold or not;but infeasible to derive any key information about the registered template as well as the query templates. We will formally define the security against both attacks in Section \ref{sec:Security-Analysis}.

\section{Proposed Threshold Predicate Encryption Scheme \label{sec:Threshold-Predicate-Encryption}}

A user-centric privacy-preserving biometric authentication scheme
requires that an end-user is able to encrypt her registered biometric
template as well as the freshly generated query templates. For
the service provider, given two encrypted templates, it should be
able to determine the distance between the two templates and compare
the distance with a threshold. In this section, we introduce Threshold
Predicate Encryption (TPE) that can fulfill the functionalities required
by such a biometric authentication system.

\subsection{Framework}

Our proposed privacy-preserving biometric authentication scheme is based on the new primitive named Threshold Predicate Encryption (TPE). Generally speaking, TPE  can be regarded as an instance of functional encryption \cite{lewko2010fully,boneh2011functional}, where decryption will output a function of the plaintext instead of the plaintext itself. The framework of functional encryption can be briefly summarized as follows. A plaintext vector $\mbfx$ is encrypted as $C_{\mbfx}$ and a secret key associated with a vector $\mbfy$ is generated as $S_{\mbfy}$. Given $C_{\mbfx}$ and $S_{\mbfy}$, the decryption will give the value of $f(\mbfx,\mbfy)$, where $f$ is a pre-defined function. Two notable instances of functional encryption are Inner Product Encryption (IPE)\cite{kim2016function}  and Predicate Encryption (PE) \cite{katz2008predicate}. The function $f$ in IPE is the inner product. That is, the decryption of IPE will give the inner product of $\mbfx$ and $\mbfy$. In comparison, PE will produce a meaningful decryption result (e.g., a flag number $0$) if and only if the inner product of $\mbfx$ and $\mbfy$ is $0$. Otherwise, the decryption result is just some random number. An important predicate is that the inner product of $\mbfx$ and $\mbfy$ equals $0$. Based on this, an extension of PE can implement exact threshold predicate encryption, meaning that the decryption result is meaningful only if the inner product of $\mbfx$ and $\mbfy$ is equal to a pre-defined threshold $\theta$.

At the high-level view, functional encryption aims at revealing only limited information about the plaintext. As introduced above, IPE reveals the inner product of the plaintext and a vector. PE reveals whether the inner product is equal to $0$ (or a threshold) or not. In application scenarios like biometric authentication, the amount of information revealed by IPE and PE are both inappropriate. As shown in our latter analysis, the inner product of $\mbfx$ and $\mbfy$ can be modeled as the distance between the registered template and the query template. As a result, IPE will give the exact distance between the two templates, which exposes too much information. With PE, one can decide whether the distance of the two templates is equal to a certain threshold, which is not sufficient for authentication purpose. What we need  is  an functional encryption scheme that can determine whether the distance between the two templates is within a threshold or not. Specifically, a TPE is composed of five algorithms:
\begin{itemize}
\item $\mathsf{TPE.Setup}()\to param$: the set up algorithm generates system parameters $param$.

\item $\mathsf{TPE.KeyGen}(\lambda)\to sk$: on input of a security parameter $\lambda$, the key generation algorithm will generate a secret key $sk$.

\item $\mathsf{TPE.Enc}(sk,\mbfx)\to C_{\mbfx}$: given a vector $\mbfx$ and the secret key $sk$, the encryption algorithm will encrypt $\mbfx$ as ciphertext $C_{\mbfx}$.

\item $\mathsf{TPE.TokenGen}(sk,\mbfy)\to T_{\mbfy}$: given a vector $\mbfy$ and the secret key $sk$, the token generation algorithm will generate a token $T_{\mbfy}$ for $\mbfy$.

\item $\mathsf{TPE.Dec}(C_{\mbfx},T_{\mbfy})\to \Lambda=\{0,1\}$: given the ciphertext $C_{\mbfx}$ and the token $T_{\mbfy}$, the decryption algorithm will output a result $\Lambda$ satisfying
\[
\Lambda=\begin{cases}
1, &  \mbfx \circ \mbfy \leq\theta\\
0, & \text{otherwise},
\end{cases}
\]
where $\mbfx \circ \mbfy$ is the inner product of $\mbfx$ and $\mbfy$.
\end{itemize}

\subsection{Design of TPE}

While our proposed TPE scheme utilizes some similar techniques as the biometric identification scheme in \cite{wang2015cloudbi}, the settings of biometric authentication are fundamentally different. In particular, our proposed TPE is designed to address the following challenges. 

\paragraph*{Challenge 1} 
The system and threat model of outsourced biometric identification and biometric authentication are different. In biometric identification, the database owner possesses the encryption and decryption keys. The aim of the server is to identify the template closest to the query template. Then the database owner will retrieve the template, decrypt it and compare the distance to a threshold. However, in our scenario, the server does not possess the decryption key thus is unable to decrypt the encrypted template and calculate the distance. What we need is an encryption scheme that can directly determine whether the distance between the query template and the registered template is within the threshold based only on ciphertexts.

\paragraph*{Challenge 2} 
The computation involved in biometric identification and authentication are different. In biometric identification, the sever needs to compute and compare the distances between a query template and all the templates in the database. However, in biometric authentication, we need to compute the distance and compare it with a threshold.

\paragraph*{Challenge 3}  The decryption process in \cite{wang2015cloudbi} will output a randomized distance between a query template and registered template. From this randomized distance, it is not easy to directly compare it with a threshold without first recovering the actual distance.

To address the above challenges, we first embed the threshold into the registered templates. To enhance security, we pad the templates with one-time randomness in a special manner and make random permutation to both the query template and registered template. After all these transformations, the decryption process can derive $\mathsf{dist}(\mathbf{x},\mathbf{y}) - \theta$, where $\mathsf{dist}(\mathbf{x},\mathbf{y})$ denotes the distance between a registered template $\mathbf{x}$ and a query template $\mathbf{y}$. 
However, if we output this value directly, it is inevitable that the exact value of  $\mathsf{dist}(\mathbf{x},\mathbf{y})$ will be exposed. Therefore, we introduce more one-time randomness into the encrypted templates. As a result, the decryption result becomes $\alpha \beta (\mathsf{dist}(\mathbf{x},\mathbf{y}) - \theta)$, where $\alpha$ and $\beta$ are positive one-time random numbers associated with $\mathbf{x}$ and $\mathbf{y}$, respectively. This design reveals only adequate information to determine whether the distance between is within the threshold and at the same time conceals the exact value of the distance.

\subsection{Construction of TPE}

Follow the aforementioned design of our threshold predicate encryption scheme, we give a detailed implementation in Protocol \ref{alg:TPE}. 
\begin{algorithm}[tbh]
\floatname{algorithm}{Protocol} 
\caption{Threshold Predicate Encryption (TPE) Scheme}{\label{alg:TPE}}

\smallskip 
\textbf{Input:} $\mbfx =\{x_1,\dots,x_n\},\mbfy =\{y_1,\dots,y_n\},\theta$.\\
\textbf{Output:} $\Lambda=\{0,1\}$.

\smallskip 
$\mathsf{TPE.Setup()}\rightarrow param$:
\begin{algorithmic}[1] 
\STATE set $param = \{ n,\theta\}$.
\end{algorithmic} 

\smallskip 
$\mathsf{TPE.KeyGen}(\lambda)\rightarrow sk $:
\begin{algorithmic}[1] 
\STATE Randomly generate two non-singular $(n+3)\times (n+3)$ matrices $M_1$ and $M_2$ and calculate their inversions $M_1^{-1}$ and $M_2^{-1}$.
\STATE Choose a random permutation $\pi\colon \mathbb{R}^{n+3} \rightarrow \mathbb{R}^{n+3}$
\STATE Set $sk =\{M_1,M_2,M_1^{-1},M_2^{-1}, \pi\}$.

\end{algorithmic} 

\smallskip
$\mathsf{TPE.Enc}(sk,\mbfx)\rightarrow C_{\mbfx}$:
\begin{algorithmic}[1]
\STATE Generate two random number $\beta$ and $r_x$.
\STATE Extend the vector $\mbfx$ to an $(n+3)$-dimensional vector $\mbfx' = (\beta x_1,\beta  x_2,\dots,\beta x_n,- \beta \theta, r_x,0)$.
\STATE Permute $\mathbf{x}'$ to obtain $\mathbf{x}'' = \pi(\mathbf{x}')$.
\STATE Transform $\mathbf{x}''$ to a diagonal matrices $X$ with $\mbox{diag}(X)=\mathbf{x}''$. 
\STATE Generate a random $(n+3)\times (n+3)$ lower  triangular matrices $S_x$ with the diagonal entries fixed as $1$. 
\STATE Compute $C_x = M_1 S_x X M_2$.
\end{algorithmic} 

\smallskip
$\mathsf{TPE.TokenGen}(sk,\mbfy)\rightarrow T_{\mbfy}$:
\begin{algorithmic}[1] 
\STATE Generate two random numbers $\alpha$ and $r_y$.
\STATE Extend $\mbfy$ to an $(n+3)$-dimensional vector $\mbfy' = (\alpha y_1,\alpha y_2,\dots,\alpha y_n,\alpha,0,r_y)$. 
\STATE Permute $\mathbf{y}'$ to obtain $\mathbf{y}'' = \pi(\mathbf{y}')$.
\STATE Transform $\mbfy''$ to a diagonal matrix $Y$ with $\mbfy''$ being the diagonal. 
\STATE Generate a random $(n+3)\times (n+3)$ lower  triangular matrix $S_y$ with the diagonal entries fixed as $1$. 

\STATE Compute $T_\mbfy = M_2^{-1}YS_y M_1^{-1}$.
\end{algorithmic} 

\smallskip
$\mathsf{TPE.Dec}(C_x,T_y)\rightarrow\Lambda=\{0,1\}$:
\begin{algorithmic}[1] 
\STATE Compute $I = \mathsf{Tr}(C_x T_y)$, where $\mathsf{Tr}(\cdot)$ denotes the trace of a matrix.

\STATE Set $\Lambda = 1$ if $I \leq 0$; otherwise set $\Lambda = 0$.
\end{algorithmic} 
\end{algorithm}

Now, we prove the correctness of the proposed TPE scheme. For a square
matrix $Y$, the trace $\mathsf{Tr}(Y)$ is defined as the sum of
the diagonal entries of $Y$. Given an invertible matrix $M_1$ of the same size, the transformation $M_1YM_1^{-1}$ is called similarity transformation of $Y$. We have the following lemma from linear algebra. 
\begin{lemma}\label{lem:trace}
The trace of a square matrix remains unchanged under similarity transformation. That is, $\mathsf{Tr}(Y)=\mathsf{Tr}(M_1YM_1^{-1})$.
\end{lemma}
Based on Lemma \ref{lem:trace}, we have the following theorem.
\begin{theorem}\label{thm:correctness}
For the proposed TPE scheme in Protocol 1,  $\Lambda \leftarrow \mathsf{TPE.Dec}(C_x,T_y)$ equals $1$ if and only if $\mathbf{x}\circ \mathbf{y} \leq \theta$, where $\mathbf{x}\circ \mathbf{y}$ denotes the inner product of $\mathbf{x}$ and $\mathbf{y}$.
\end{theorem}

\begin{IEEEproof}
Following the procedure in Protocol \ref{alg:TPE}, the vector $\mbfx$ is transformed to $C_x = M_1S_x XM_2$. The vector $\mbfy$ is transformed to $T_y=M_2^{-1}YS_yM_1^{-1}$. Then we have $C_xT_y=M_1S_xXYS_yM_1^{-1}$. From Lemma \ref{lem:trace}, we have 
$I=\mathsf{Tr}(C_x T_y)=\mathsf{Tr}(S_xXYS_y)$.
Since $S_x$ and  $S_y$  are selected as lower triangular matrices, where all the diagonal entries are set to $1$, the diagonal entries of $S_xX$ and $YS_y$ are all the same as those of $X$ and $Y$. Thus we have $I=\mathsf{Tr}(XY)$.
Since $X$ and $Y$ are diagonal matrices, $I=\mathbf{x}''\circ\mbfy''=\mbfx'\circ\mbfy'=\alpha\beta(\mbfx\circ\mbfy-\theta)$.
Since $\alpha$ and $\beta$ are positive, we have $\Lambda=1$ (i.e.,$I\leq 0$) if and only if $\mbfx\circ\mbfy \leq\theta$.

\vskip-1.0\baselineskip
\end{IEEEproof}

\section{Biometric Authentication Under Different Distance Metrics\label{sec:Biometric-Authentication}}

In this section, we will first introduce some necessary background on biometric authentication. Then, we show how to construct privacy-preserving biometric authentication systems utilizing our proposed TPE scheme under different distance metrics.

\subsection{Backgrounds}\label{sec-backgrounds}
The first critical step in biometric authentication is to efficiently transform biometric traits into templates that are easy for computation. Such a process is often called \emph{feature extraction}. The extracted features are often represented as \emph{feature vectors}. Depending on the biometric traits, the process as well as the result of feature extraction could differ. For example, a fingerprint can be transformed to a FingerCode \cite{jain1999fingercode,jain2000filterbank,jain1999multichannel} that is a vector of integers with dimension $640$. An Iris image is often represented as a binary string of $2048$ bits. In the following, we briefly review the feature extraction process of fingerprints. The details can be found in \cite{jain2000filterbank,jain1999multichannel}.

\begin{figure}[h]
\centering
\includegraphics[width=\columnwidth]{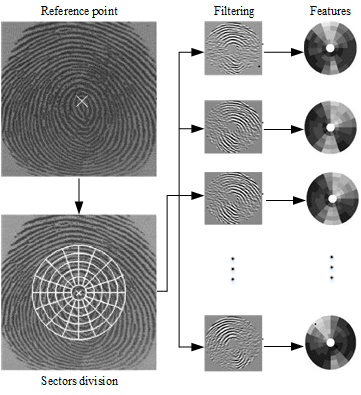}
\caption{Feature extraction of fingerprints: (i) Identify reference point; (ii) Divide region of interest into sectors around reference point; (iii) Filter region of interest; (iv) Extract features. }
\label{fig-finger}
\end{figure}

As illustrated in Fig. \ref{fig-finger}\footnote{This figure is partially obtained from \cite{jain1999multichannel}.}, given an image of a fingerprint, the first step is to identify a reference point. Then the region of interest around the reference point is divided into $5$ bands and $16$ sectors. Those sectors are further normalized and filtered by $8$ different Gabor filters. At last, the features are extracted from each filtered image. The final result is a $640$-dimensional vector (FingerCode) representing each fingerprint image, where each entry in the vector is an $8$-bit integer. An import feature of the FingerCode is that it is translation invariant, meaning that translation of the fingerprint image would not result in much difference in the FingerCode. However, FingerCode is not rotation invariant. As a result, rotation of images will often cause different FingerCodes. To resolve this issue, a user is often associated with several (for example, 5)  FingerCodes captured from rotated images in the database. In the following discussion, we assume that at the local side, there exists a sensor that can capture the end-user's biometric trait and transform it to a multi-dimensional vector. 

In a user-centric biometric authentication system, an end-user will
send her encrypted biometric template to the service provider in the
registration phase. In the query phase, the end-user will encrypt a freshly generated template and send it to the service provider for authentication usage. Thus, a critical issue is to decide whether two templates are close enough. These problem is reduced to measuring the distance of two vectors in a metric space and compare the distance to a certain threshold. Such a compute-then-compare computational model on encrypted data is well suited for our proposed TPE scheme. 

Furthermore, different biometric templates often rely on different similarity measurements. For example, in Iris recognition, the templates are represented by binary vectors and the similarity is generally measured by Hamming distance. For fingerprint, the Euclidean distance is normally utilized to measure the similarity. Our proposed TPE scheme is highly flexible in that it can be applied to measuring similarity based on different distance metrics. As a result, TPE can be utilized as the critical component to build different privacy-preserving biometric authentication systems. In the rest of this section, we will illustrate how to utilize TPE to construct a biometric authentication scheme based on Euclidean distance, Hamming distance and so on.

\subsection{Euclidean Distance}

Euclidean distance is often used to measure the similarity between
vectors of non-binary entries. A
FingerCode representing a fingerprint is an $n$-dimensional vector,
where each entry is an $l$-bit integer. Typically, $n=640$ and $l=8$.
We denote a registered FingerCode as $\mbfx=(x_1,x_2,\dots,x_n)$
and a query FingerCode as $\mbfy=(y_1,y_2,\dots,y_n)$. Let $\mathsf{d}_E(\mbfx,\mbfy)$ be the Euclidean distance between $\mbfx$ and $\mbfy$. Then we have
\[
\mathsf{d}_E^2(\mbfx,\mbfy)=\sum_{i=1}^nx_i^2+\sum_{i=1}^ny_i^2-2\mbfx\circ\mbfy,
\]
where $\mbfx\circ\mbfy$ is the inner product of $\mbfx$
and $\mbfy$. Let $\theta$ be a pre-defined threshold.
Our goal is to extend $\mbfx$ and $\mbfy$ 
to  vectors $\mbfx'$ and $\mbfy'$ respectively such that the relation $\mathsf{d}_E^2(\mbfx,\mbfy)< \theta^2$ can be determined through computing $\mbfx' \circ \mbfy'$. In light of
this, we let $\mbfx'=(2x_1,2x_2,\dots,2x_n,-\sum\limits_{i=1}^nx_i^2,1,\theta^2)$
and $\mbfy'=(y_1,y_2,\dots,y_n,1,-\sum\limits_{i=1}^ny_i^2,1)$.
Then we have 
\begin{eqnarray*}
\mbfx'\circ\mbfy' & = & 2\mbfx\circ\mbfy+\theta^2-\sum_{i=1}^nx_i^2-\sum_{i=1}^ny_i^2\\
 & = & \theta^2-\mathsf{d}_E^2(\mbfx,\mbfy).
\end{eqnarray*}
To secure the biometric templates, we further add different randomnesses (i.e., $\alpha,\beta,r_x$ and $r_y$) to the extended vectors as shown in Protocol \ref{pro:bioAuth}. The rest of the encryption procedures is then the same as those in $\mathsf{TPE.Enc}$ and $\mathsf{TPE.TokenGen}$.

As presented in Protocol \ref{pro:bioAuth}, during the registration phase, an end-user encrypts his template $\mathbf{x}$ as $C_x$ and registers $C_x$ along with her identity at the service provider. During the query phase, the end-user encrypts a freshly generated template $\mathbf{y}$ as $T_y$ and sends $T_y$ to the service provider. Then the service provider runs $\mathsf{TPE.Dec}$ with inputs $C_x$ and $T_y$ and outputs an authentication result. The correctness of this scheme is guaranteed by Theorem \ref{thm:correctness}, with slight adaption to Euclidean distance. That is $\Lambda = \mathsf{Authenticated}$ if and only if $\mathsf{d}_E (\mathbf{x},\mathbf{y}) \leq \theta$.
\begin{algorithm}[!htp] 
\floatname{algorithm}{Protocol} 
\caption{Privacy Preserving Biometric Authentication}{\label{pro:bioAuth}}

\smallskip 
\textbf{Input:} $\mbfx =\{x_1,\dots,x_n\},\mbfy =\{y_1,\dots,y_n\},\theta$.\\
\textbf{Output:} $\Lambda=\{\mathsf{Denied},\mathsf{Authenticated}\}$.

\smallskip 
\textbf{Setup} (End-user $U$):
\begin{algorithmic}[1] 
\STATE Set the public parameters as $param = \{ n,\theta\}$.
\STATE Randomly generate two matrices $M_1$ and $M_2$ with dimension $(n+5)\times (n+5)$ and a permutation $\pi: \mathbb{R}^{n+5} \rightarrow \mathbb{R}^{n+5}$. 
\STATE Set secret key $sk =\{M_1,M_2,M_1^{-1},M_2^{-1},\pi\}$.
\end{algorithmic} 

\smallskip
\textbf{Registration} (End-user $U$):
\begin{algorithmic}[1] 
\STATE Generate  random numbers $\beta$ and $r_x$. Eextend $\mbfx$ to an $(n+5)$-dimensional vector $\mbfx' = (2\beta x_1,2\beta x_2,\dots,2\beta x_n,-\beta \sum\limits_{i=1}^nx_i^2,\beta ,\beta \theta^2,r_x,0)$.
\STATE Permute $\mathbf{x}'$ to obtain $\mathbf{x}'' = \pi(\mathbf{x}')$.
\STATE Transform $\mathbf{x}''$ to a diagonal matrices $X$ with  $\mathbf{x}''$ being the diagonal.
\STATE Generate a random  $(n+5)\times (n+5)$ lower triangular matrix $S_x$ with the diagonal entries fixed as $1$. Compute $C_x = M_1 S_x X M_2$.
\STATE Register the record $\langle ID_U,C_x\rangle$ to the service provider $SP$, where $ID_U$ is the identity of end-user $U$.
\end{algorithmic} 

\smallskip
\textbf{Query} (End-user $U$):
\begin{algorithmic}[1] 
\STATE Generate random numbers $\alpha$ and $r_y$.
\STATE Extend $\mbfy$ to an $(n+5)$-dimensional vector $\mbfy' = (\alpha y_1, \alpha y_2,\dots,\alpha y_n,\alpha,-\alpha \sum\limits_{i=1}^ny_i^2,\alpha, 0,r_y)$.
\STATE Permute $\mathbf{y}'$ to obtain $\mathbf{y}'' = \pi(\mathbf{y}')$.
\STATE Transform $\mbfy''$ to a diagonal matrix $Y$ with diagonal being $\mathsf{y}''$.
\STATE Generate a random $(n+5)\times (n+5)$ lower triangular matrix $S_y$ with the diagonal entries fixed as $1$. Compute $T_y = M_2^{-1} Y S_y  M_1^{-1}$
\STATE Send the query $\langle ID_U, T_y\rangle$ to $SP$.
\end{algorithmic} 

\smallskip
\textbf{Authentication} (Service Provider $SP$):

\begin{algorithmic}[1] 
\STATE On receiving a query from the end-user $U$, retrieve the registered record according to $ID_U$. 
\STATE Compute $I = \mathsf{Tr}(C_x T_y)$. 
\STATE Set $\Lambda = \mathsf{Authenticated}$ if $I \geq 0$; otherwise set $\Lambda = \mathsf{Denied}$.
\end{algorithmic} 
\end{algorithm}

\subsection{Distance in Hamming Space\label{subsec:Hamming-Distance}}

From the construction of Euclidean distance, we know that the
critical part in computing the distance through inner product lies
in proper design of the extended vectors. Thus, in the following, we will focus on 
how to design the vectors in order to compute different distances. 

Hamming distance is a popular metric to measure the similarity of
binary template such as Iris. Now, we assume the registered template
and query template are $\mbfx=(x_1,x_2,\dots,x_n)$ and
$\mbfy=(y_1,y_2,\dots,y_n)$ respectively, where $x_i$
and $y_i$ are $0$ or $1$. To calculate the Hamming distance $\mathsf{d}_H(\mbfx,\mbfy)$ between
$\mbfx$ and $\mbfy$, we first map
the $0$'s in $\mbfx$ and $\mbfy$ to $-1$ and map $1$'s
to $1$. Then we have
\[
2\mathsf{d}_H(\mbfx,\mbfy)=n-\mbfx\circ\mbfy.
\]
The condition $\mathsf{d}_H(\mbfx,\mbfy)-\theta \leq 0$
is equivalent to $\mbfx\circ\mbfy + 2\theta -n \geq 0$. Thus, we need to design vectors $\mathbf{x}'$ and $\mathbf{y}'$ such that $\mbfx\circ\mbfy + 2\theta -n$ can be represented as $\mathbf{x}' \circ \mathbf{y}'$. In light of this, we let $\mbfx'=(\beta x_1,\beta x_2,\dots,\beta x_n,\beta(2\theta-n),r_x,0)$
and $\mbfy'=(\alpha y_1,\alpha y_2,\dots,\alpha y_n,\alpha,0,r_y)$. Then the rest of the authentication process is similarly as in Protocol \ref{pro:bioAuth}. 

In fact, the Hamming distance between two binary vectors is just one specific distance metric. There are many other different  metrics such as Minkowski distance, Sokal \& Michener
similarity and Sokal \& Sneath-II ~\cite{choi2010survey} introduced for different applications. Using our proposed TPE scheme, we are able to evaluate such metrics and compare them to a pre-defined threshold. The critical part is to properly design the vectors $\mathbf{x}'$ and $\mathbf{y}'$ given two binary vectors $\mathbf{x}$ and $\mathbf{y}$.

\section{Security Analysis\label{sec:Security-Analysis}}

In this section, we analyze the security of PassBio under both passive
attack and active attack as defined in Section \ref{sec:Problem-Statement}.
PassBio is designed so that the service provider
is unable to learn any critical information about the registered and query templates other than what is already revealed by
the decryption process, given an encrypted registered template
and a sequence of encrypted query templates.

Since PassBio is based on our proposed TPE, we will focus on the security analysis of TPE in the following discussion. An important difference
between TPE and some traditional symmetric encryption schemes is that
it is the service provider (could be malicious) that carries out
the decryption process. And the decryption process will reveal whether the inner product is within a threshold or not. Therefore, in the security analysis of TPE, it is necessary to analyze the security of both the encryption and decryption process, which will be discussed separately in the following sections. 

\subsection{Encryption Security\label{subsec:Encryption-Security}}

We first give a sketch of encryption security analysis. We will first
utilize two experiments to model the ability of the adversary in passive
attack and active attack, respectively. Then, we define the security
of TPE under both passive and active attacks. At last, we prove
the security of TPE under active attack since it implies the security under passive attack.

\subsubsection{Security against passive attack}

In our scenario, the passive attack corresponds to the ciphertext-only-attack \cite{katz2014introduction}, where an adversary $\mathcal{A}$
observes a sequence of ciphertext. We define an experiment $\ensuremath{\mathsf{Passive}_{\mathcal{A},\mathsf{TPE}}^{\mathsf{mult}}(\lambda)}$
to simulate passive attacks, where the superscript $\mathsf{mult}$
denotes that the adversary $\mathcal{A}$ is able to submit multiply
messages instead of one single message. 
\begin{algorithm}
\caption*{\textbf{Passive attack experiment}  $\mathsf{Passive}_{\mathcal{A},\mathsf{TPE}}^{\mathsf{mult}}(\lambda)$:}

\smallskip 
\begin{algorithmic}[1] 
\STATE Given a security parameter $\lambda$, the adversary $\mathcal{A}$ outputs two sequences of messages $M_0 = (m_{0,1},m_{0,2},\dots,m_{0,t})$ and $M_1 = (m_{1,1},m_{1,2},\dots,m_{1,t})$, where the length of each message $| m_{0,i}| = |m_{1,i}|, i=1,2,\dots,t$.

\STATE The challenger $\mathcal{C}$ runs $\mathsf{TPE.KeyGen}(\lambda)$ to generate the secret key $sk$. 

\STATE $\mathcal{C}$ chooses a uniform bit $b\in \{0,1\}$ and computes the ciphertext $c_i=\mathsf{TPE.TokenGen}(m_{b,i},sk)$. The sequence $C =(c_1,c_2,\dots,c_t)$ is returned to $\mathcal{A}$.

\STATE The adversary $\mathcal{A}$ outputs a bit $b'$.

\STATE The output of the experiment is 1 if $b=b'$, and 0 otherwise.
\end{algorithmic}
\end{algorithm}

Based on $\ensuremath{\mathsf{Passive}_{\mathcal{A},\mathsf{TPE}}^{\mathsf{mult}}(\lambda)}$,
we now define the security of TPE under passive attack.
\begin{definition}
The proposed TPE scheme is secure against passive attack if for all
polynomial-time adversary $\mathcal{A}$, there is a negligible function
$\mathsf{negl}$ such that the probability
\[
|\mathsf{Pr}(\mathsf{Passive}_{\mathcal{A},\mathsf{TPE}}^{\mathsf{mult}}(\lambda)=1)-\frac{1}{2}| \leq \mathsf{negl}(\lambda).
\]
\end{definition}

\begin{remark}
In the above security definition, we only use the token generation
function $\mathsf{TPE.TokenGen}$ as a representative. This is because
the operations involved in $\mathsf{TPE.Enc}$ and $\mathsf{TPE.TokenGen}$
are almost the same. The security analysis for $\mathsf{TPE.Token}$
applies for $\mathsf{TPE.Enc}$. However, in our security proof, we
will show that both $\mathsf{TPE.Enc}$ and $\mathsf{TPE.TokenGen}$
meet the security requirement. 
\end{remark}

Based on Definition 1, we have the following theorem.
\begin{theorem}\label{thm:passive}
The proposed TPE scheme is secure against passive attack.
\end{theorem}
We will omit the proof of Theorem \ref{thm:passive}. Instead, we
will prove security against active attack since it implies the security under passive attack.

\subsubsection{Security against active attack}

Under the active attack, the service provider is able to observe a
sequence of pairs of query templates as well as their encrypted version.
This can happen when, for example, some adversaries submit their templates
and collude with the service provider. This attack scenario corresponds
to the Chosen-Plaintext-Attack (CPA) in cryptography. Accordingly,
an encryption scheme has CPA-security if it is secure against CPA.
To prove that TPE has CPA-security, we model the active attack using
and experiment $\ensuremath{\mathsf{Active}_{\mathcal{A},\mathsf{TPE}}(\lambda)}$. 
We define CPA-security for TPE as follows.

\begin{algorithm}
	\caption*{\textbf{Active attack experiment} $\mathsf{Active}_{\mathcal{A},\mathsf{TPE}}(\lambda)$:}
	\smallskip 
	\begin{algorithmic}[1] 
		\STATE The function $\mathsf{TPE.KeyGen}(\lambda)$ generates a secret key $sk$.
		
		\STATE The adversary $\mathcal{A}$ is given oracle access to the function $\mathsf{TPE.TokenGen}(sk,\cdot)$ and  outputs two messages $m_0$ and $m_1$ of the same length to the challenger $\mathcal{C}$.
		
		\STATE The challenger $\mathcal{C}$ chooses a uniform bit $b\in\{0,1\}$, then computes $c=\mathsf{TPE.TokenGen}(sk,m_b)$ and returns to $\mathcal{A}$.
		
		\STATE $\mathcal{A}$ continues to have oracle access to $\mathsf{TPE.TokenGen}(sk,\cdot)$ and outputs a bit $b'$. Note however, $\mathcal{A}$ cannot use $\mathsf{TPE.TokenGen}(sk,\cdot)$ to generate tokens for messages somehow related to $m_0$ and $m_1$.
		\STATE The output of the experiment is 1 if $b=b'$, and 0 otherwise.
	\end{algorithmic}
\end{algorithm} 

\begin{definition}
The proposed TPE is secure against active attack if for all polynomial-time
adversary $\mathcal{A}$, there is a negligible function $\mathsf{negl}$
such that the probability
\[
|\mathsf{Pr}(\mathsf{Active}_{\mathcal{A},\mathsf{TPE}}(\lambda)=1)- \frac{1}{2}| \leq\mathsf{negl}(\lambda).
\]
\end{definition}

\begin{remark}
Different from the passive attack experiment, the adversary will continually
have oracle access to the token generation function. This models the
situation where the adversary is able to observe multiple pairs of messages
and their ciphertexts. 
\end{remark}
\begin{remark}
Unlike the passive attack experiment where the adversary submits multiple
pairs of messages, we only discuss the situation where the adversary
submits one pair of messages $(m_0,m_1)$ to the challenger. This
is because it is proved in \cite{katz2014introduction} that any private-key
encryption scheme that is CPA-secure is also CPA-secure for multiple
encryptions. As a result, it is sufficient to prove that TPE is CPA-secure
for one single encryption.
\end{remark}
\begin{theorem}\label{thm:active}
The proposed TPE is secure against active attack.
\end{theorem}

\begin{IEEEproof}
We need to prove that the adversary $\mathcal{A}$ cannot distinguish
$\mathsf{TPE.TokenGen}(sk,m_0)$ and $\mathsf{TPE.TokenGen}(sk,m_1)$,
even given the oracle access to $\mathsf{TPE.TokenGen}(sk,\cdot)$. 

Consider the encryption of message $m_0$. Suppose $m_0=(m_{0,1},m_{0,2},\dots,m_{0,n})$
is an $n$-dimensional vector. Follow the procedure in $\mathsf{TPE.TokenGen}$,
the vector $m_0$ is first extended to a vector $m_0'=(\alpha m_{0,1},\alpha m_{0,2},\dots,\alpha m_{0,n},\alpha,0,r_0)$,
where $\alpha$ and $r_0$ are random numbers. The vector $m_0'$ is then permuted as $m_0''$, which 
is then extended to an $(n+3)\times(n+3)$ diagonal matrix $Y_0$.
Then, the ciphertext for $m_0$ is $c_0=M_2^{-1}Y_0S_0M_1^{-1}$, where $S_0$ is a random lower triangular matrix. We note that
the product of $Y_0$ and $S_0$ will produce a lower triangular
matrix denoted as $G_0$, with $m_0'$ as the diagonal. Now we
focus on the product $c_0=M_2^{-1}G_0M_1^{-1}$. 

Denote the entries in $M_2^{-1}$ and $M_1^{-1}$ as $a_{ij}$
and $b_{ij}$, respectively, where $i,j=1,2,\dots,n+3$. For matrix
$G_0$, denote its non-zero entries in the lower triangular part
as $s_{ij}$, where $i>j$ and $i,j=1,2,\dots,n+3$. Then, by law
of matrix multiplication, each entry $c_{ij}$ in $c_0$ can be
written in the form of 
\begin{eqnarray}
c_{ij} & = & \sum [f_{ij}^1(a_{ij},b_{ij})m_i +f_{ij}^2(a_{ij},b_{ij})\alpha\nonumber\\
 &  & +f_{ij}^{3}(a_{ij},b_{ij})r_0 +f_{ij}^{4}(a_{ij},b_{ij},s_{ij})],
\label{eq:entry}
\end{eqnarray}
where $f_{ij}^{t}$, $t=1,2,3,4$ are polynomials. Equation (\ref{eq:entry})
is obtained by summing up each terms of $m_i$, $\alpha$ and $r_0$,
respectively. 


Now, observe Equation (\ref{eq:entry}) in the context of the experiment
$\mathsf{Active}_{\mathcal{A},\mathsf{TPE}}(\lambda)$. We know that $a_{ij}$
and $b_{ij}$ are fixed. $a, r$ and $s_{ij}$ are one-time random
numbers. $m_i$ are chosen and can be controlled by the adversary
$\mathcal{A}$. In step 4) of experiment $\mathsf{Active}_{\mathcal{A},\mathsf{TPE}}(\lambda)$,
the adversary $\mathcal{A}$ can select different $m_i$ each time
and observe the value of $c_{ij}$ since $\mathcal{A}$ continuously
has oracle access to $\mathsf{TPE.TokenGen}(sk_i,\cdot)$. However,
since $a,r$ and $s_{ij}$ are one-time random numbers, the polynomials
$f_{ij}^2(a_{ij},b_{ij})\alpha$, $f_{ij}^{3}(a_{ij},b_{ij})r$
and $f_{ij}^{4}(a_{ij},b_{ij},s_{ij})$ all looks random to $\mathcal{A}$.
As a result, the summation $c_{ij}$ looks random to $\mathcal{A}$.
This means that, for any message $m$ chosen by $\mathcal{A}$ and
its corresponding ciphertext, $\mathcal{A}$ cannot distinguish which
message is actually encrypted. Thus, the adversary $\mathcal{A}$
can only output $b'$ by randomly guessing. Thus we have 
\[
|\mathsf{Pr}(\mathsf{Active}_{\mathcal{A},\mathsf{TPE}}(\lambda)=1)-\frac{1}{2}| \leq \mathsf{negl}(\lambda).
\]
\vskip-1.5\baselineskip
\end{IEEEproof}

\subsection{Decryption Security}

The decryption function $\mathsf{TPE.Dec}$ outputs an intermediate
result denoted as $R = C_x T_y$  and a final result $I=\mathsf{Tr}(R)$.
In the following security analysis, we discuss what information can
be learned by the service provider from $R$ and $I$. 

As in Protocol \ref{alg:TPE}, $R=M_1S_xXYS_yM_1^{-1}$,
where $S_x$ and $S_y$ are random matrices. Recall the proof for Theorem \ref{thm:active},
where $c_0=M_2^{-1}G_0M_1^{-1}$. Since matrix $G_0$ and
$XY$ follow the same construction, it is obvious that the transformation $R=M_1S_x XYS_yM_1^{-1}$ also has CPA-security. In other words, the transformation is semantically secure, meaning that the adversary is not able to derive any key information of $X$ and $Y$ from $R$.

Now, for the final result $I=\alpha\beta(\mbfx\circ\mbfy-\theta)$,
we define a decryption oracle $\mathcal{O}$ as follows.
\begin{algorithm}
\caption*{\textbf{Decryption Oracle} $\mathcal{O}$:}

\smallskip 
\begin{algorithmic}[1] 
\STATE The oracle $\mathcal{O}$ fixes a vector $\mbfx$ and a number $\theta$.
\STATE For any submitted vector $\mbfy$, $\mathcal{O}$ generates two positive random numbers $\alpha$ and $\beta$ and output $\gamma = \alpha \beta (\mbfx \circ \mbfy-\theta)$.
\end{algorithmic}
\end{algorithm}

\begin{theorem}
The oracle $\mathcal{O}$ does not have CPA-security. \label{thm:oracle}
\end{theorem}

\begin{IEEEproof}
We provide a proof sketch since the CPA-security proof process follows
that for Theorem \ref{thm:active}.

An adversary $\mathcal{A}$ is able to continuously have access to
$\mathcal{O}$. $\mathcal{A}$ will submit $\mbfy_i$ at her
own choice and observe the output $\gamma_i$. Since $\alpha$ and
$\beta$ are positive, it is possible that there exists $\mbfy_1$
and $\mbfy_2$ such that $\gamma_1>0$ while $\gamma_2<0$.
This means that, in an experiment defined for CPA-security, the adversary
$\mathcal{A}$ is able to distinguish two ciphertext for two submitted
messages. By definition, the oracle $\mathcal{O}$ does not have CPA-security.
\end{IEEEproof}
Theorem \ref{thm:oracle} states that the final result $I$ actually
reveals some information about $\mbfx$ and $\mbfy$. This
result is expected in our design since we want to determine if the inner
product of $\mbfx$ and $\mbfy$ is within a threshold $\theta$
or not from the final result $I$. However, we note that in our proposed
TPE, every vector $\mbfy$ is associated with a one-time independent
random number $\alpha$ and every vector $\mbfx$ is associated
with a one-time random number $\beta$. As a result, in the active
attack, what an adversary can observe through decryption is a series
of results $I_i=\alpha_i\beta(\mbfx\circ\mbfy_i-\theta)$.
Since $\alpha_i$ are selected independently, the final results
$I_i$ only reveals whether $\alpha_i\beta(\mbfx\circ\mbfy-\theta)$
is positive or not. No more key information can be derive from $I_i$.

\subsection{The Effect of Randomness on Security\label{subsec:The-Effect-of}}

Besides the randomly generated long-time keys (i.e., $M_1$,
$M_2$ and $\pi$), we also introduce different one-time randomness in the
encryption scheme. At the high-level view, the one-time randomness provides
TPE with CPA-security similar to that of the one-time pad. From a cryptographic
point of view, the one-time pad encryption scheme provides perfect security.
However, it is not practical since the one-time secret key has the
same length as the message itself. The most notable difference between
TPE and the traditional encryption schemes is that TPE actually does not
decrypt the message. Instead, TPE evaluate a function of the
ciphertext in order to obtain the function value of the plaintext.
As a result, TPE does not require the one-time randomness in the decryption
process. In this sense, TPE can achieve  the security comparable to the  one-time
pad while avoiding the impractical key management requirement.

It is important to understand the effect of different randomness on
security. We briefly categorize the one-time randomness utilized by
TPE into three types.
\begin{itemize}
\item Type I: \textit{result-disguising randomness}. When extending the vectors in  both $\mathsf{TPE.Enc}$ and $\mathsf{TPE.TokenGen}$, we
use random $\beta$ and $\alpha$ respectively to multiply with each
entry of $\mbfx$ and $\mbfy$. Since $\alpha$ and $\beta$
will remain in the decryption result, we name it as result-disguising
randomness.
\item Type II: \textit{vector-extension randomness}. In both $\mathsf{TPE.Enc}$
and $\mathsf{TPE.TokenGen}$,
we extend the vector and pad it with a random $r$.
\item Type III: \textit{matrix-multiplication randomness}. In both $\mathsf{TPE.Enc}$
and $\mathsf{TPE.TokenGen}$, we multiply the extended matrices ($X$ and $Y$) with random matrices ($S_x$ and $S_y$).
\end{itemize}
These one-time randomnesses together ensure the CPA-security of the
encryption process of TPE as analyzed in Section \ref{subsec:Encryption-Security}.
The main function of decryption is to evaluate the trace of the matrix. We note that the trace function will cancel Type II and Type III randomness. However, Type I randomness
will remain in the decryption result.
This is important since it will only reveal partial information of
the plaintext, which is just adequate for the purpose of biometric authentication. We will further demonstrate the effect of Type I randomness in Section \ref{subsec:Improved-Security-for}.

\section{Other Applications of TPE}\label{sec:Applications-of-Threshold}

Our proposed threshold predicate encryption scheme enables a compute-then-compare
computational model over encryption data. That is, given two encrypted
vector $\mbfx$ and $\mbfy$, an untrusted party is able
to determine whether the inner product of $\mbfx$ and $\mbfy$
is greater than or within a threshold $\theta$. No other key information
about the value of $\mbfx, \mbfy$ or $\mbfx\circ\mbfy$
is exposed. Previously, we also showed that utilizing the inner product
of $\mbfx$ and $\mbfy$, we are able to compute many distance
and similarity metrics. Such properties of TPE are critical for many applications
that require data security and privacy.

\subsection{Improved Security for Outsourced Biometric Identification\label{subsec:Improved-Security-for}}

Outsourcing of different computational problems to the cloud while preserving the security and privacy of the outsourced problem has becoming a new trend. Many previous works have considered  secure
outsourcing of different problems \cite{zhou2016linsos,zhou2016secure,zhou2017expsos,zhou2016ABE,yuan2013efficient,wang2015cloudbi}.
In \cite{wang2015cloudbi}, a secure outsourcing scheme is proposed
for biometric identification. The system models of outsourced biometric
identification and biometric authentication are fundamentally different.
In outsourced biometric identification, a data owner possesses a database
of users' biometric templates. The goal of biometric identification
is that given a query template, the data owner needs to identify a
user to whom the query template belongs to. 

To protect the security and privacy of biometric templates, \cite{wang2015cloudbi}
proposed an outsourcing scheme where the database owner will first
encrypt the templates and then outsource the encrypted data to the
cloud. Specifically, the data owner encrypts a biometric template
$\mbfx$ as $C_x$ using a symmetric key $sk$. For a given
query template $\mathbf{z}$, it is also encrypted as $C_{z}$ using
the same key $sk$. The scheme is designed in such a manner that given
two encrypted templates $C_x$ and $C_y$ and a query template
$C_{z}$, the cloud is able to determine which template ($\mbfx$
or $\mbfy$) is closer to $\mathbf{z}$, without learning any
key information about $\mbfx$, $\mathbf{z}$ and $\mbfy$.
By repeating this process, the cloud is able to identify the template
$\mbfx$ that is closest to $\mathbf{z}$. Then the encrypted
version $C_x$ is returned to the data owner, who can decrypt $C_x$
to obtain $\mbfx$ and calculate the actual distance between
$\mbfx$ and $\mathbf{z}$. Thus, the data owner can finally
decide whether $\mbfx$ and $\mbfy$ are close enough such
that they belong to the same person. 

There are mainly two security and privacy issues regarding the above
scheme. First, the registration phase is vulnerable to the registration attack \cite{hahn2016poster}, since an adversary (i.e., the cloud) is able to inject known templates into the database. During decryption,
the cloud is able to derive the following equation (i.e., Equation
(3) in \cite{hahn2016poster}):
\[
b_{ci}=\frac{(\mathsf{Tr}(Y_i^{'}B_{c}^{'})-\mathsf{Tr}(X_i'B_{c}^{'}))-(y_{i(n+1)}-x_{i(n+1)})}{y_{ii}-x_{ii}},
\]
 where $b_{ci}$ is the $i$-th entry in a submitted query template
$\mbfb_{c}$. Since $\mathsf{Tr}(Y_i^{'}B_{c}^{'})$ and $\mathsf{Tr}(X_i'B_{c}^{'})$
are computable and $\mbfx$ and $\mbfy$ are selected by
the cloud, the cloud is able to recover $b_{ci}$. Repeating such
attack will finally recover the whole query template $\mbfb_{c}$
as demonstrated in \cite{hahn2016poster}. 

Second, from the decryption result, the cloud is able to
learn more information than needed. In particular, the cloud is able to determine which one of any two encrypted
template is closer to the query template. By repeating this process,
the cloud can actually rank all the templates by their distances to
the query template. This unnecessarily reveals more information than
what is needed in biometric identification.

We now show that our proposed TPE scheme can address these two issues.
 The security vulnerability of the scheme in \cite{wang2015cloudbi} was caused due to lacking of Type I randomness as defined in Section \ref{subsec:The-Effect-of}.
The trace function $\mathsf{Tr}(\cdot)$ will cancel the Type III randomness, resulting in  Equation (3) in \cite{hahn2016poster}.

Our proposed TPE scheme can be directly utilized in
outsourced biometric identification. In the encryption part, each
registered template $\mbfx$ is  encrypted
with $\mathsf{TPE.Enc}$. A query template $\mathbf{z}$ is encrypted
with $\mathsf{TPE.TokenGen}$. The decryption process will give $\alpha_{z}\beta_x(\mathsf{dist}^2(\mbfx,\mathbf{z})-\theta^2)$,
where $\alpha_{z}$ and $\beta_x$ are one-time random numbers associated
with $\mathbf{z}$ and $\mbfx$ respectively. As a result, Equation
(3) in \cite{hahn2016poster} is replaced by 
\[
b_{ci}=\frac{(\mathsf{Tr}(PB_{c}^{'})-\mathsf{Tr}(QB_{c}^{'}))-(p_{n+1}-q_{n+1})}{\alpha_{c}\beta_x(p_n-q_n)}.
\]
 Note that $\alpha_{c}$ is a one-time random number associated with
a query $b_{c}$ and $\beta_x$ is a one-time random number associated
with $\mbfx$. Thus, although the adversary is able to insert
known templates into the database, it cannot derive $b_{ci}$ due
to the one-time randomness. In other words, the outsourced biometric
identification scheme based on TPE is able to defend against registration
attack.

For the second privacy issue, the decryption result $\alpha_{z}\beta_x(\mathsf{dist}^2(\mbfx,\mathbf{z})-\theta)$
will only reveal whether the distance between the query $\mathbf{z}$
and the registered template $\mbfx$ is within a threshold or not.
Since $\beta_x$ is a one-time randomness associated with each registered
template $\mbfx$, the relative distance information is concealed.
As a result, the cloud is not able to rank all the registered templates
according to the distance to the query template.

\subsection{Searching Over Encrypted Data}

With the development of cloud computing and storage, there is a clear
motivation for searching over encrypted data \cite{song2000practical,shi2007multi,boneh2007conjunctive,tu2013processing}.
For example, a medical institution may store its medical data in the
cloud. To ensure security of the data, the institution chooses
to encrypt all the data before outsourcing. Meanwhile, the institution
wishes to maintain the searching ability over the encrypted data in
order to retrieve the desired data files. The proposed TPE is a promising
solution for searching over encrypted data. In the following, we discuss
how to utilize TPE to implement different searching functionalities.

\subsubsection{Set Intersection}

We assume that a file $F_i$ is indexed by a set of keywords $S_i$.
The files and their associated keyword sets are encrypted and outsourced
to the cloud. A search query consists of a set of keywords $S_{j}$.
Given the search query, the cloud will return the file $F_i$ if
the overlap of keyword sets $S_i$ and $S_{j}$ exceeds a certain
threshold $\theta$. That is $|S_i\cap S_{j}|>\theta$. 

The above set intersection search function can be implemented through
TPE as follows. Suppose the universe of keywords is the set $S$ with
size $n$. Fix the order of the keywords within $S$. Then, an index
$S_i$ can be formulated as an $n$-dimensional binary vector $\mbfx_i$,
where $x_i^{t}=1$ means that the $t$-th keyword in $S$ appears
in $S_i$. The vector $\mbfx_i$ for file $F_i$ is encrypted
using $\mathsf{TPE.Enc}$. Each file is then encrypted using standard
symmetric encryption schemes such as AES. The encrypted files and
index are outsourced to the cloud. For a search query $S_{j}$, a
vector $\mbfx_{j}$ can be formulated in a similar manner. Then
a search token can be generated using $\mathsf{TPE.TokenGen}$. With
this formulation, it is obvious that $|S_i\cap S_{j}|=\mbfx_i\circ\mbfx_{j},$
where $\mbfx_i\circ\mbfx_{j}$ denotes the inner product
of $\mbfx_i$ and $\mbfx_{j}$. With TPE, the cloud is
able to identify the files whose associated indices satisfy $\mbfx_i\circ\mbfx_{j}>\theta$
while not learning any useful information about the indices.

\subsubsection{Weighted Sum Evaluation}

For many numeric data, it is significant to evaluate the weighted
sum of the data record with different weights. For example, the grades
of each subject for a student form a vector $G_i$. An evaluator
wants to evaluate the performance of the students via some criteria.
Each criterion can be formulated as the weighed sum of the grades.
The different weights reflects different emphasis on the subjects. 

We assume that an administrator possess the grades for all the students.
For privacy issues, all the grades are encrypted using $\mathsf{TPE.Enc}$
and stored in an external server. An evaluator desires to identify
those students whose performance meets certain standard. In this scenario,
the evaluator can submit a vector of weights $W_{j}$ to the administrator,
who will then generate a search token for the evaluator through $\mathsf{TPE.TokenGen}$.
The evaluator can submit the token generated for $W_{j}$ to the sever
and search over the encrypted grades. The server is then able to identify
the students whose grades satisfy $G_i\circ W_{j}>\theta$. 

\section{Performance Evaluation\label{sec:Performance-Evaluation}}

In this section, we evaluate the performance of PassBio. First, we
give detailed analysis of both computational and communication complexity.
Then, some numeric results are presented for the proposed TPE through
simulation. 

 \subsection{Complexity Analysis}

As shown in Protocol \ref{pro:bioAuth}, at local side an end-user
needs to run the $\mathsf{TPE.KeyGen}$, $\mathsf{TPE.Enc}$ and $\mathsf{TPE.TokenGen}$
algorithms. The service provider needs to run the $\mathsf{TPE.Dec}$
algorithm for every query. It is obvious that the computational bottleneck
of these algorithms lies in matrix multiplication or matrix inversion.
Thus, in the following analysis, we will focus on matrix multiplication
and inversion. Without loss of generality, we assume that the matrices
involved in the computation all have the same dimension $n\times n$. 

For the function $\mathsf{TPE.KeyGen}$, two random matrices are generated
and two matrix inversions need to be calculated. Note that the setup phase is generally a one-time process. That is, $\mathsf{TPE.KeyGen}$ 
needs to be executed by the end-user only once. The function $\mathsf{TPE.Enc}$ and $\mathsf{TPE.TokenGen}$ will both take $3$ matrix multiplications. As a result, they have a complexity of $\mathcal{O}(n^3)$, without optimization for matrix multiplication.

In the function $\mathsf{TPE.Dec}$, the trace of $C_xT_y$ needs to be computed. There is no need to calculate the matrix multiplication before evaluating the trace. Only computing of the diagonal entries is needed. Thus, $\mathsf{TPE.Dec}$ has a complexity of $\mathcal{O}(n^2)$.

In terms of communication overhead, assume all the matrix or vector has the same size $l$. In the registration phase, the end-user needs to submit the encrypted template $C_x$
to the service provider. Thus the communication overhead for registration
is $n^2l$. Similarly, the communication overhead for the query
phase is also $n^2l$.

\subsection{Efficiency Improvement}

The above complexity analysis shows that the computational bottleneck of both $\mathsf{TPE.Enc}$ and $\mathsf{TPE.TokenGen}$ lie in matrix multiplication. For resource-constrained devices such as mobile phones, the computation of matrix multiplication with high dimensions is still expensive, if not impossible. In the following, we will introduce two typical techniques that can reduce the computational overhead for mobile devices.

\subsubsection{Dimension Reduction}\label{sec-dimension}

The complexity of normal matrix multiplication is $\mathcal{O}(n^3)$, where $n$ is the dimension of the matrices. Thus, a straight forward way to reduce the complexity is to reduce the dimension of the matrices. For applications such as biometric authentication and identification, it is critical to preserve the identification accuracy while reducing the dimension. Several works \cite{bianchi2010implementing,kambhatla2006dimension,mandal2009curvelet} have been devoted to reducing the sizes of biometric templates. In~\cite{bianchi2010implementing}, two techniques are introduced to decimate the FingerCode representation. The \emph{tesselation reduction} approach reduces the dimension of FingerCode from the feature generation phase, which is illustrated in Section \ref{sec-backgrounds}. Specifically, given a fingerprint image, this approach will reduce the number of sectors of the tessellation. The other approach is to directly apply some general dimension reduction methods such as PCA to the obtained FingerCodes. In this way, the most compact representation of FingerCode is found for a specific dataset.

We note that the above two approaches will both degrade the identification accuracy, however, to a satisfying level. In the experiments~\cite{bianchi2010implementing}, the length of FingerCode vary from 640 to 8 in the tesselation approach. For PCA approach, the dimension of FingerCode varies from 64 to 4. Generally speaking, the shorter the FingerCode is, the worse the accuracy would be. However, the experimental result demonstrated that FingerCode of dimensions 96 (from tesselation reduction) and 8 (from PCA) can achieve a satisfactory accuracy compared to that of the original 640. We also note that the approaches in \cite{bianchi2010implementing} quantized each entry in FingerCode resulting a reduced accuracy. However, our proposed TPE scheme can be directly utilized to real numbers. Thus, TPE is applicable to the non-quantized case in \cite{bianchi2010implementing}, which has a higher accuracy.

\subsubsection{Online/Offline Computation}\label{sec-online}

The idea of online/offline computation \cite{chow2011identity,hohenberger2014online,liu2009efficient} is to divide a computational expensive process into an online phase and an offline phase. During the offline phase, some pre-computation is done without given the input. During the online phase, given the input, it is relatively easy to padding the offline computation result in order to generate the final result. Typically, the offline computation is carried out when the mobile devices are idle or getting charged. Thus, such an approach can reduce the overall responding time and battery consumption.

Our proposed TPE scheme can utilize such approach to reduce the online computational overhead. For example, in the query phase, an end-user needs to compute $M_2^{-1} Y S_y M_1 ^{-1}$ given a transformed template $Y$. Then during the offline phase, the end-user can generate the random matrix $S_y$ and compute $S_y M_1^{-1}$. The computation results can be stored for later usage. When a fresh template $Y$ is generated, the end-user can compute $M_2^{-1} Y S_y M_1 ^{-1}$ during the online phase. This approach can reduce half of the computational overhead, which is critical for resource-constrained devices.

\subsection{Numeric Results}

In this section, we measure the performance of our proposed TPE scheme through simulation. Since the functions $\mathsf{TPE.Setup}$ and $\mathsf{TPE.KeyGen}$ are both one-time processes during the registration phase, we mainly focus on the execution time of $\mathsf{TPE.TokenGen}$.

Since PassBio is a user-centric biometric authentication scheme, we measure the performance on both mobile phone and personal laptop. In the simulation, we utilize a mobile phone with Android 6.0 operating system, 2.5 GHz Cortex-A72 CPU and 4 GB RAM. We also utilize a personal laptop with macOS 10, 1.6 GHz Intel Core i5 and 4 GB RAM. The java library UJMP \cite{UJMP} and C++ library Armadillo are utilized for the simulation in the mobile phone and personal computer, respectively. We note that the performance relies on the selection of software packages. Our selection does not guarantee the best performance. Through complexity analysis, we know that the most important parameter affecting the performance is the dimension $n$ of the vector. For the simulation on the mobile phone and laptop, we let $n$ vary from 10 to 300 and from 100 to 2000, respectively. Due to the dimension reduction techniques introduced in Section \ref{sec-dimension}, the dimension $n=300$ is sufficient for most of the biometric templates. We also utilize the online/offline computation mechanism introduced in Section \ref{sec-online} to reduce to online computational overhead. 
\begin{figure}[h]
	\centering
	\includegraphics[width=0.4\textwidth]{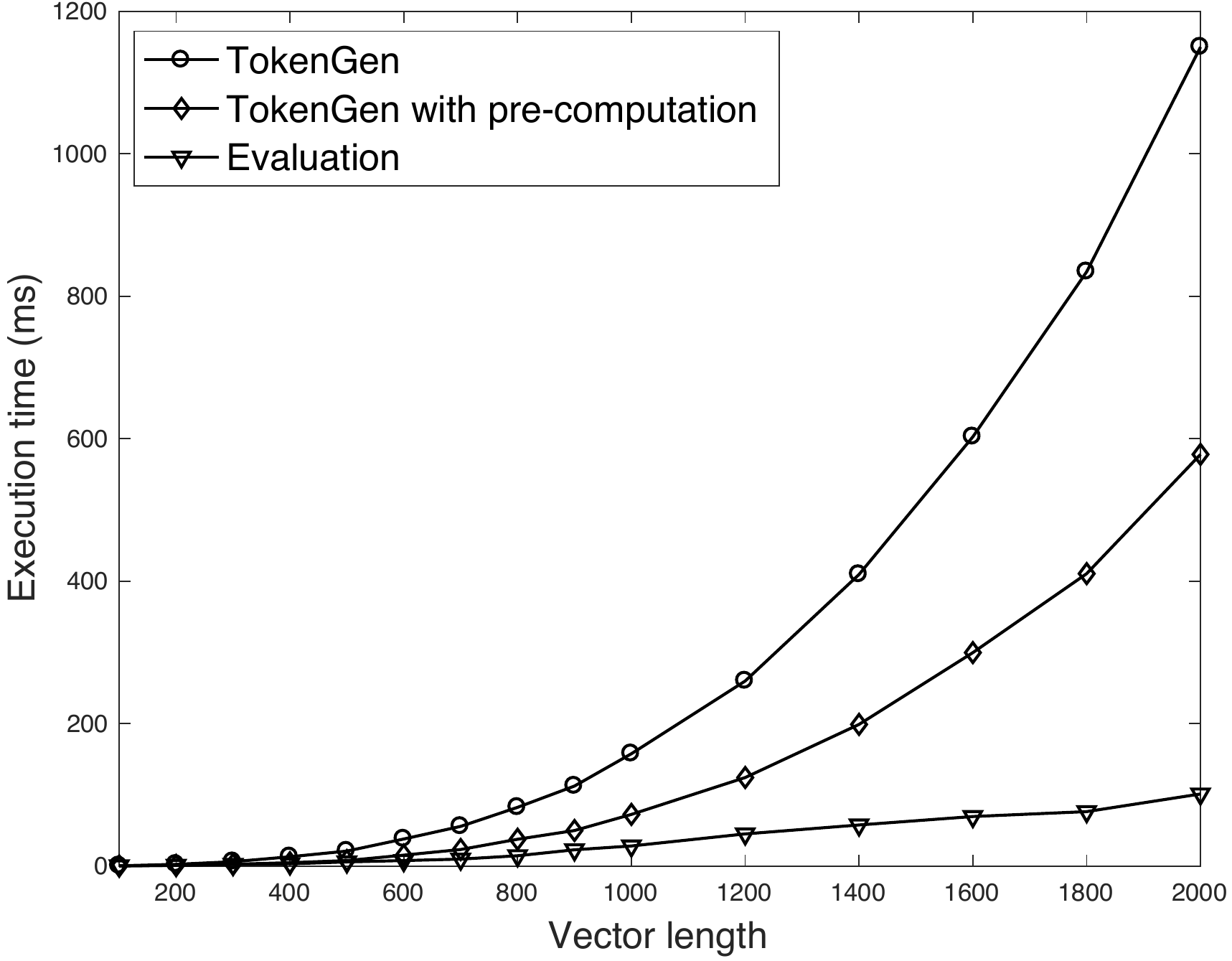}
	\caption{Performance of token generation and evaluation simulated on laptop (with vs. without pre-computation)}
	\label{sim-computer}
\end{figure}

\begin{figure}[h]
	\centering
	\includegraphics[width=0.4\textwidth]{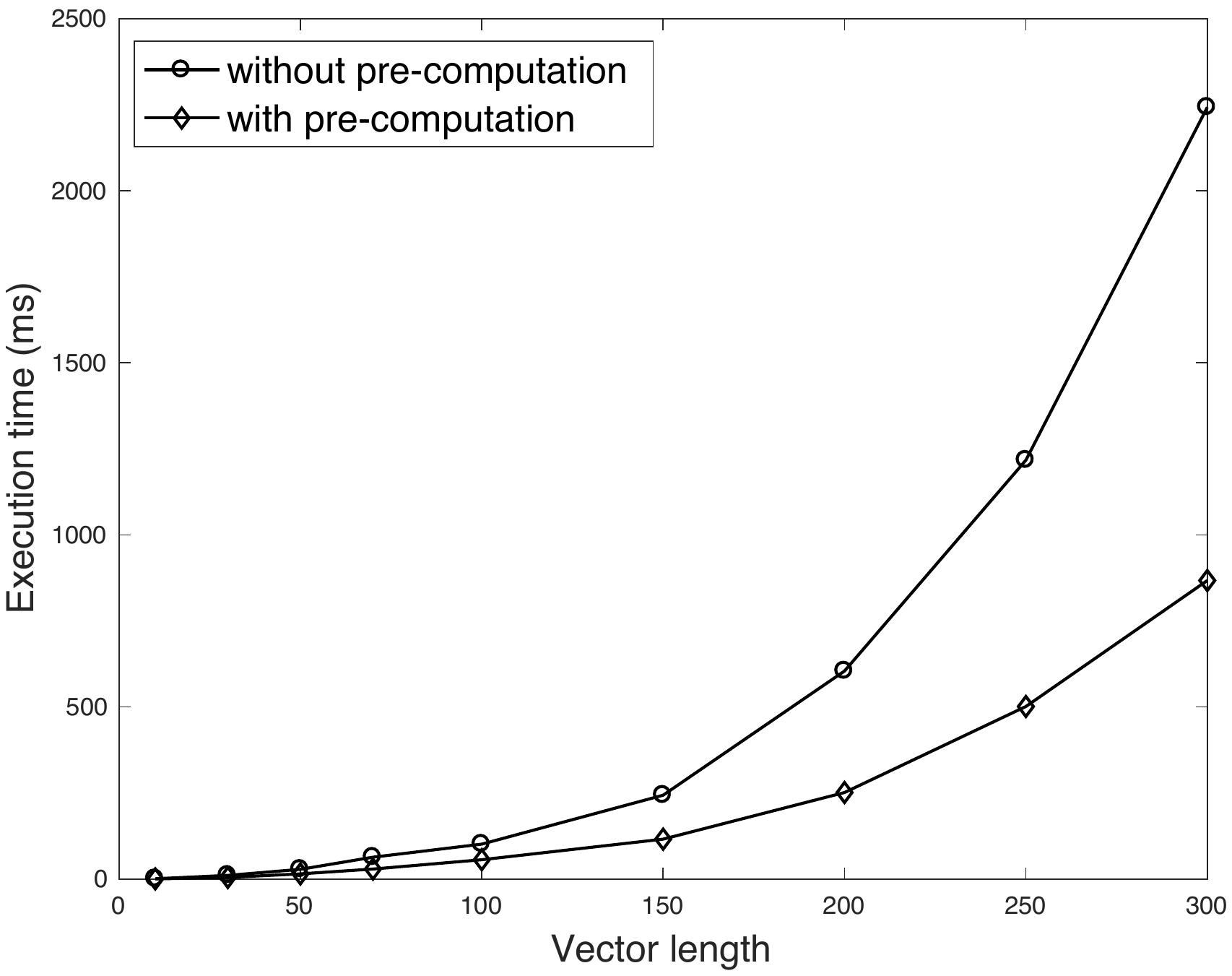}
	\caption{Performance of token generation and evaluation on mobile phone (with vs. without pre-computation)}
	\label{sim-phone}
\end{figure}

The numeric result on the laptop is shown in Fig. \ref{sim-computer}. The token generation time for moderate size template ($n$ is around 200) is just around one millisecond with pre-computation. For high-dimensional template with $n=2000$, the token generation time is less than 1 second with pre-computation. The numeric result on the mobile phone is shown in Fig. \ref{sim-phone}. The simulation results show that it is efficient to generate tokens for templates with moderate size. For example, when $n=100$, the generation time is approximately 50 $ms$. When $n=300$, the generation time is around 900 $ms$. It can be observed in both figures that the online/offline mechanism can effectively reduce the online computational overhead. By pre-computation during the offline phase, the online computation time is reduced to about half of the whole processing time.

\section{Conclusion\label{sec:Conclusion}}

In this paper, we proposed a Threshold Predicate Encryption~(TPE) scheme. TPE is able to encrypt a vector $\mathbf{x}$ and generate a token for another vector $\mathbf{y}$. Given the two encrypted vectors, any party is able to determine whether the inner product of $\mathbf{x}$ and $\mathbf{y}$ is within a pre-defined threshold or not. Our security analysis shows that no sensitive information about the vectors can be learned by the untrusted party under both passive and active attacks. Based on TPE, we proposed PassBio, a privacy preserving user-centric biometric authentication scheme. One key feature of PassBio is that end-users can encrypt their own biometric template and register it to the service provider. Then the end-user is able to encrypt their freshly generated template and submit them to the service provider for authentication usage. We show that the TPE is suitable for a compute-then-compare computational model on encrypted data. Such a computational model can be widely used in many applications requiring computations on encrypted data while preserving the data security and privacy. In particular, we presented two additional applications of TPE, searching over encrypted data and outsourced biometric identification. Our simulation results demonstrated that the proposed TPE can be efficiently implemented on both mobile phones and personal laptops.

\bibliographystyle{plain}
\bibliography{citation}

\begin{thebibliography}{10}

\bibitem{UJMP}
Universal java matrix package.
\newblock \url{https://ujmp.org/}.

\bibitem{abdalla2015simple}
Michel Abdalla, Florian Bourse, Angelo De~Caro, and David Pointcheval.
\newblock Simple functional encryption schemes for inner products.
\newblock In {\em IACR International Workshop on Public Key Cryptography},
  pages 733--751. Springer, 2015.

\bibitem{bianchi2010implementing}
Tiziano Bianchi, Stefano Turchi, Alessandro Piva, Ruggero~Donida Labati,
  Vincenzo Piuri, and Fabio Scotti.
\newblock Implementing fingercode-based identity matching in the encrypted
  domain.
\newblock In {\em Biometric Measurements and Systems for Security and Medical
  Applications (BIOMS), 2010 IEEE Workshop on}, pages 15--21. IEEE, 2010.

\bibitem{bishop2015function}
Allison Bishop, Abhishek Jain, and Lucas Kowalczyk.
\newblock Function-hiding inner product encryption.
\newblock In {\em International Conference on the Theory and Application of
  Cryptology and Information Security}, pages 470--491. Springer, 2015.

\bibitem{boneh2011functional}
Dan Boneh, Amit Sahai, and Brent Waters.
\newblock Functional encryption: Definitions and challenges.
\newblock {\em Theory of Cryptography}, pages 253--273, 2011.

\bibitem{boneh2007conjunctive}
Dan Boneh and Brent Waters.
\newblock Conjunctive, subset, and range queries on encrypted data.
\newblock {\em Theory of cryptography}, pages 535--554, 2007.

\bibitem{choi2010survey}
Seung-Seok Choi, Sung-Hyuk Cha, and Charles~C Tappert.
\newblock A survey of binary similarity and distance measures.
\newblock {\em Journal of Systemics, Cybernetics and Informatics}, 8(1):43--48,
  2010.

\bibitem{choi2014secure}
Sunoh Choi, Gabriel Ghinita, Hyo-Sang Lim, and Elisa Bertino.
\newblock Secure knn query processing in untrusted cloud environments.
\newblock {\em IEEE Transactions on Knowledge and Data Engineering},
  26(11):2818--2831, 2014.

\bibitem{chow2011identity}
Sherman~SM Chow, Joseph~K Liu, and Jianying Zhou.
\newblock Identity-based online/offline key encapsulation and encryption.
\newblock In {\em Proceedings of the 6th ACM Symposium on Information, Computer
  and Communications Security}, pages 52--60. ACM, 2011.

\bibitem{datta2016functional}
Pratish Datta, Ratna Dutta, and Sourav Mukhopadhyay.
\newblock Functional encryption for inner product with full function privacy.
\newblock In {\em Public-Key Cryptography--PKC 2016}, pages 164--195. Springer,
  2016.

\bibitem{elmehdwi2014secure}
Yousef Elmehdwi, Bharath~K Samanthula, and Wei Jiang.
\newblock Secure k-nearest neighbor query over encrypted data in outsourced
  environments.
\newblock In {\em Data Engineering (ICDE), 2014 IEEE 30th International
  Conference on}, pages 664--675. IEEE, 2014.

\bibitem{goldwasser2014multi}
Shafi Goldwasser, S~Dov Gordon, Vipul Goyal, Abhishek Jain, Jonathan Katz,
  Feng-Hao Liu, Amit Sahai, Elaine Shi, and Hong-Sheng Zhou.
\newblock Multi-input functional encryption.
\newblock In {\em Annual International Conference on the Theory and
  Applications of Cryptographic Techniques}, pages 578--602. Springer, 2014.

\bibitem{hahn2016poster}
Changhee Hahn and Junbeom Hur.
\newblock Poster: Towards privacy-preserving biometric identification in cloud
  computing.
\newblock In {\em Proceedings of the 2016 ACM SIGSAC Conference on Computer and
  Communications Security}, pages 1826--1828. ACM, 2016.

\bibitem{hohenberger2014online}
Susan Hohenberger and Brent Waters.
\newblock Online/offline attribute-based encryption.
\newblock In {\em International Workshop on Public Key Cryptography}, pages
  293--310. Springer, 2014.

\bibitem{jain2012biometric}
Anil~K Jain and Karthik Nandakumar.
\newblock Biometric authentication: System security and user privacy.
\newblock {\em IEEE Computer}, 45(11):87--92, 2012.

\bibitem{jain201650}
Anil~K Jain, Karthik Nandakumar, and Arun Ross.
\newblock 50 years of biometric research: Accomplishments, challenges, and
  opportunities.
\newblock {\em Pattern Recognition Letters}, 79:80--105, 2016.

\bibitem{jain1999multichannel}
Anil~K Jain, Salil Prabhakar, and Lin Hong.
\newblock A multichannel approach to fingerprint classification.
\newblock {\em IEEE transactions on pattern analysis and machine intelligence},
  21(4):348--359, 1999.

\bibitem{jain1999fingercode}
Anil~K Jain, Salil Prabhakar, Lin Hong, and Sharath Pankanti.
\newblock Fingercode: a filterbank for fingerprint representation and matching.
\newblock In {\em Computer Vision and Pattern Recognition, 1999. IEEE Computer
  Society Conference on.}, volume~2, pages 187--193. IEEE, 1999.

\bibitem{jain2000filterbank}
Anil~K Jain, Salil Prabhakar, Lin Hong, and Sharath Pankanti.
\newblock Filterbank-based fingerprint matching.
\newblock {\em IEEE transactions on Image Processing}, 9(5):846--859, 2000.

\bibitem{kambhatla2006dimension}
Nandakishore Kambhatla and Todd~K Leen.
\newblock Dimension reduction by local principal component analysis.
\newblock {\em Dimension}, 9(7), 2006.

\bibitem{katz2014introduction}
Jonathan Katz and Yehuda Lindell.
\newblock {\em Introduction to modern cryptography}.
\newblock CRC press, 2014.

\bibitem{katz2008predicate}
Jonathan Katz, Amit Sahai, and Brent Waters.
\newblock Predicate encryption supporting disjunctions, polynomial equations,
  and inner products.
\newblock {\em Advances in Cryptology--EUROCRYPT 2008}, pages 146--162, 2008.

\bibitem{kim2016function}
Sam Kim, Kevin Lewi, Avradip Mandal, Hart~William Montgomery, Arnab Roy, and
  David~J Wu.
\newblock Function-hiding inner product encryption is practical.
\newblock {\em IACR Cryptology ePrint Archive}, 2016:440, 2016.

\bibitem{lewko2010fully}
Allison~B Lewko, Tatsuaki Okamoto, Amit Sahai, Katsuyuki Takashima, and Brent
  Waters.
\newblock Fully secure functional encryption: Attribute-based encryption and
  (hierarchical) inner product encryption.
\newblock In {\em Eurocrypt}, volume 6110, pages 62--91. Springer, 2010.

\bibitem{liu2009efficient}
Joseph~K Liu and Jianying Zhou.
\newblock An efficient identity-based online/offline encryption scheme.
\newblock In {\em ACNS}, volume 5536, pages 156--167. Springer, 2009.

\bibitem{mandal2009curvelet}
Tanaya Mandal, QM~Jonathan Wu, and Yuan Yuan.
\newblock Curvelet based face recognition via dimension reduction.
\newblock {\em Signal Processing}, 89(12):2345--2353, 2009.

\bibitem{naehrig2011can}
Michael Naehrig, Kristin Lauter, and Vinod Vaikuntanathan.
\newblock Can homomorphic encryption be practical?
\newblock In {\em Proceedings of the 3rd ACM workshop on Cloud computing
  security workshop}, pages 113--124. ACM, 2011.

\bibitem{rane2013secure}
Shantanu Rane, Ye~Wang, Stark~C Draper, and Prakash Ishwar.
\newblock Secure biometrics: concepts, authentication architectures, and
  challenges.
\newblock {\em IEEE Signal Processing Magazine}, 30(5):51--64, 2013.

\bibitem{shen2009predicate}
Emily Shen, Elaine Shi, and Brent Waters.
\newblock Predicate privacy in encryption systems.
\newblock In {\em TCC}, volume 5444, pages 457--473. Springer, 2009.

\bibitem{shi2007multi}
Elaine Shi, John Bethencourt, TH~Hubert Chan, Dawn Song, and Adrian Perrig.
\newblock Multi-dimensional range query over encrypted data.
\newblock In {\em Security and Privacy, 2007. SP'07. IEEE Symposium on}, pages
  350--364. IEEE, 2007.

\bibitem{smart2010fully}
Nigel~P Smart and Frederik Vercauteren.
\newblock Fully homomorphic encryption with relatively small key and ciphertext
  sizes.
\newblock In {\em Public Key Cryptography}, volume 6056, pages 420--443.
  Springer, 2010.

\bibitem{song2000practical}
Dawn~Xiaoding Song, David Wagner, and Adrian Perrig.
\newblock Practical techniques for searches on encrypted data.
\newblock In {\em Security and Privacy, 2000. S\&P 2000. Proceedings. 2000 IEEE
  Symposium on}, pages 44--55. IEEE, 2000.

\bibitem{tu2013processing}
Stephen Tu, M~Frans Kaashoek, Samuel Madden, and Nickolai Zeldovich.
\newblock Processing analytical queries over encrypted data.
\newblock In {\em Proceedings of the VLDB Endowment}, volume~6, pages 289--300.
  VLDB Endowment, 2013.

\bibitem{wang2016practical}
Boyang Wang, Yantian Hou, and Ming Li.
\newblock Practical and secure nearest neighbor search on encrypted large-scale
  data.
\newblock In {\em Computer Communications, IEEE INFOCOM 2016-The 35th Annual
  IEEE International Conference on}, pages 1--9. IEEE, 2016.

\bibitem{wang2015cloudbi}
Qian Wang, Shengshan Hu, Kui Ren, Meiqi He, Minxin Du, and Zhibo Wang.
\newblock Cloudbi: Practical privacy-preserving outsourcing of biometric
  identification in the cloud.
\newblock In {\em European Symposium on Research in Computer Security}, pages
  186--205. Springer, 2015.

\bibitem{wong2009secure}
Wai~Kit Wong, David Wai-lok Cheung, Ben Kao, and Nikos Mamoulis.
\newblock Secure knn computation on encrypted databases.
\newblock In {\em Proceedings of the 2009 ACM SIGMOD International Conference
  on Management of data}, pages 139--152. ACM, 2009.

\bibitem{yuan2013efficient}
Jiawei Yuan and Shucheng Yu.
\newblock Efficient privacy-preserving biometric identification in cloud
  computing.
\newblock In {\em INFOCOM, 2013 Proceedings IEEE}, pages 2652--2660. IEEE,
  2013.

\bibitem{zhou2017expsos}
Kai Zhou, MH~Afifi, and Jian Ren.
\newblock Expsos: Secure and verifiable outsourcing of exponentiation
  operations for mobile cloud computing.
\newblock {\em IEEE Transactions on Information Forensics and Security},
  12(11):2518--2531, 2017.

\bibitem{zhou2016linsos}
Kai Zhou and Jian Ren.
\newblock Linsos: Secure outsourcing of linear computations based on affine
  mapping.
\newblock In {\em Communications (ICC), 2016 IEEE International Conference on},
  pages 1--5. IEEE, 2016.

\bibitem{zhou2016ABE}
Kai Zhou and Jian Ren.
\newblock Secure fine-grained access control of mobile user data through
  untrusted cloud.
\newblock In {\em Computer Communication and Networks (ICCCN), 2016 25th
  International Conference on}, pages 1--9. IEEE, 2016.

\bibitem{zhou2016secure}
Kai Zhou and Jian Ren.
\newblock Secure outsourcing of scalar multiplication on elliptic curves.
\newblock In {\em Communications (ICC), 2016 IEEE International Conference on},
  pages 1--5. IEEE, 2016.

\end{thebibliography}

\end{document}